\documentclass[epjST]{svjour}
\usepackage{graphicx}
\usepackage{amsmath}

\begin{document}

\newcommand \ER {Erd\H{o}s-Reny\'i\ }
\newcommand \governancenetwork{G^g}
\newcommand \usernetwork{G^u}
\newcommand \resourcenetwork{G^r}

\newcommand \userwaitingtime{T_u}
\newcommand \userwaitingtimei{T_{u,i}}

\newcommand \governancewaitingtime{T_g}
\newcommand \governancewaitingtimei{T_{g,i}}

\newcommand \governancestate {S}

\newcommand \governanceninternallinks{L^g_{kl}}
\newcommand \userinternallinks{L^u_{ij}}

\newcommand \extractionlinks{L^{ur}_{i}}
\newcommand \harvestlinks{L^{ur}_{i}}
\newcommand \reportinglinks{L^{ug}_{ik}}
\newcommand \taxationlinks{L^{gu}_{ki}}

\newcommand \resourcenodes{V^{r}}
\newcommand \usernodes{V^{u}}
\newcommand \governancenodes{V^{g}}

\newcommand \numberofresourcenodes{N_{r}}
\newcommand \numberofusernodes{N_{u}}
\newcommand \numberofgovernancenodes{N_{g}}

\newcommand \resourcei{s_i}
\newcommand \harvesti{h_i}
\newcommand \harvestj{h_j}

\newcommand \efforti{E_i}
\newcommand \effortj{E_j}

\newcommand \taxrate{\gamma}

\newcommand \tcrit{\Delta\userwaitingtime{}_{,crit}}
\newcommand \gammacrit{\gamma{}_{crit}}

\title{The physics of governance networks:\\ critical transitions in contagion dynamics on multilayer adaptive networks with application to the sustainable use of renewable resources}

\author{Fabian Geier\inst{1,2}, Wolfram Barfuss\inst{3,4}\fnmsep\thanks{\email{barfuss@pik-potsdam.de}}, Marc Wiedermann\inst{1}\fnmsep\thanks{\email{marcwie@pik-potsdam.de}}, J\"urgen Kurths\inst{1,4,5}, Jonathan F. Donges\inst{3, 6}}

\institute{Complexity Science, Potsdam Institute for Climate Impact Research, Member of the Leibniz Association, Potsdam, Germany \and Department of Physics, Ludwig Maximilians University, Munich, Germany \and Earth System Analysis, Potsdam Institute for Climate Impact Research, Member of the Leibniz Association, Potsdam, Germany \and Department of Physics, Humboldt University, Berlin, Germany \and Saratov State University, Saratov, Russia \and Stockholm Resilience Centre, Stockholm University, Stockholm, Sweden}

\abstract{
Adaptive networks are a versatile approach to model phenomena such as contagion and spreading dynamics, critical transitions and structure formation that emerge from the dynamic coevolution of complex network structure and node states. Adaptive networks have been successfully applied to study and understand phenomena ranging from epidemic spreading, infrastructure, swarm dynamics and opinion formation to the sustainable use of renewable resources.
Here, we study critical transitions in contagion dynamics on multilayer adaptive networks with dynamic node states and present an application to the governance of sustainable resource use. We focus on a three layer adaptive network model, where a polycentric governance network interacts with a social network of resource users which in turn interacts with an ecological network of renewable resources.
We uncover that sustainability is favored for slow interaction timescales, large homophilic network adaptation rate (as long it is below the fragmentation threshold) and high taxation rates. Interestingly, we also observe a trade-off between an eco-dictatorship (reduced model with a single governance actor that always taxes unsustainable resource use) and the polycentric governance network of multiple actors. In the latter setup, sustainability is enhanced for low but hindered for high tax rates compared to the eco-dictatorship case.
These results highlight mechanisms generating emergent critical transitions in contagion dynamics on multilayer adaptive network and show how these can be understood and approximated analytically, relevant for understanding complex adaptive systems from various disciplines ranging from physics and epidemiology to sociology and global sustainability science. The paper also provides insights into potential critical intervention points for policy in the form of taxes in the governance of sustainable renewable resource use that can inform more process-detailed social-ecological modeling.} 

\maketitle

\section{Introduction}
\label{sec:level1}

Adaptive networks are a flexible approach to model phenomena such as contagion and spreading phenomena, critical transitions and structure formation that emerge from the dynamic coevolution of complex network structure and node states~\cite{holme_nonequilibrium_2006,gross_adaptive_2008,gross2009adaptive}. Adaptive networks have been successfully applied to study and understand phenomena ranging from epidemic spreading~\cite{gross_epidemic_2006} and early warning signals for critical transitions therein~\cite{horstmeyer2018network}, swarm dynamics~\cite{couzin2011uninformed,huepe2011adaptive}, evolution of autocatalytic sets~\cite{jain1998autocatalytic,jain2001model}, opinion formation~\cite{holme_nonequilibrium_2006} and spreading of behaviors such as smoking~\cite{schleussner2016clustered} to the sustainable use of renewable resources and modelling social-ecological transformations~\cite{wiedermann_macroscopic_2015,barfuss_sustainable_2017,lade2017modelling,mullerhansen2019can}. Recently, adaptive dynamics have also been studied in multilayer network models that allow for representing different types of nodes or agents and their complex interconnections in a structured way~\cite{boccaletti2014structure,amato2017opinion,min2019multilayer}.

Adaptive networks are also recognized as a promising approach to build a bridge between theoretical physics and efforts to understand future trajectories of the Earth system in the Anthropocene where human social dynamics has become a dominant geological process~\cite{verburg2016methods,donges2017closing}. By modelling complex social systems as adaptive multilayer networks embedded in land-use~\cite{arneth2014global} or more comprehensive Earth system models~\cite{muller2017towards,donges2018earth}, methods from complex systems theory, nonlinear dynamics and statistical physics can be applied to identify management options, critical transitions, tipping points and critical intervention points towards sustainable development~\cite{milkoreit2018social}, map out safe operating spaces for these systems~\cite{heitzig2016topology,mathias2018does,anderies2019knowledge}, and more generally, analyze complex co-evolutionary dynamics of human-environment systems including the evolution of technological and knowledge systems~\cite{renn2017extended,herrmann2018case}. Important recent challenges in this field include the identification of sensitive intervention points for policy~\cite{farmer2019sensitive} and adaptive multi-level governance strategies~\cite{mathias2017multi} that can help to overcome systemic blockages and initiate the deep social-ecological transformations~\cite{lade2017modelling} needed to avoid dangerous anthropogenic climate change and degradation of biosphere integrity~\cite{steffen2018trajectories}.

While the control of adaptive network dynamics has already been studied in the context of opinion formation influenced via zealotry~\cite{klamser2017zealotry}, a stylized form of lobbyism, there is an increasing interest in studying modern polycentric, adaptive and multi-level governance and management of social-ecological systems from a complex systems perspective~\cite{mathias2017multi}, creating bridges to the theory of governance networks from political science ~\cite{koliba2018governance}. In this paper, we derive and analyze an adaptive multilayer network model to investigate the dynamics of an adaptive and polycentric governance network interacting with an adaptive social network of users of private renewable resources, extending upon the recently proposed and studied copan:EXPLOIT model~\cite{wiedermann_macroscopic_2015,barfuss_sustainable_2017}. In the extended model, termed copan:TAXPLOIT in this paper, governance nodes can either penalize associated unsustainable resource users by introducing taxes or can be indifferent to the level of resource exploitation (i.e., by not introducing any tax). The trait of enforcing such an environmental tax can spread contagiously on the governance network via social learning. Additionally, the governance network can adapt via homophilic rewiring. Analogously, in the resource user layer the trait of sustainable or unsustainable resource exploitation can spread via social learning and the users' social network can adapt via homophilic rewiring as well (Fig.~\ref{fig:model_scheme}).

We study this multilayer adaptive network system using numerical simulations and analytical approximations. We particularly focus on analyzing the conditions under which adaptive polycentric governance fosters the sustainable use of renewable resources and increases the resilience and size of the sustainable safe operating space of the system. We also identify critical transitions and tipping points, e.g. in the tax rate parameter, that separate domains of sustainable and unsustainable outcomes in parameter space.

We introduce the studied multilayer adaptive network model in detail in Sect.~\ref{sec:model}, report and discuss results of numerical simulations and analytical approximations (Sect.~\ref{sec:results}), and finish with concluding remarks (Sect.~\ref{sec:conclusions}).

\section{Model description}
\label{sec:model}

\begin{figure}[t!]
	\includegraphics[width=\textwidth]{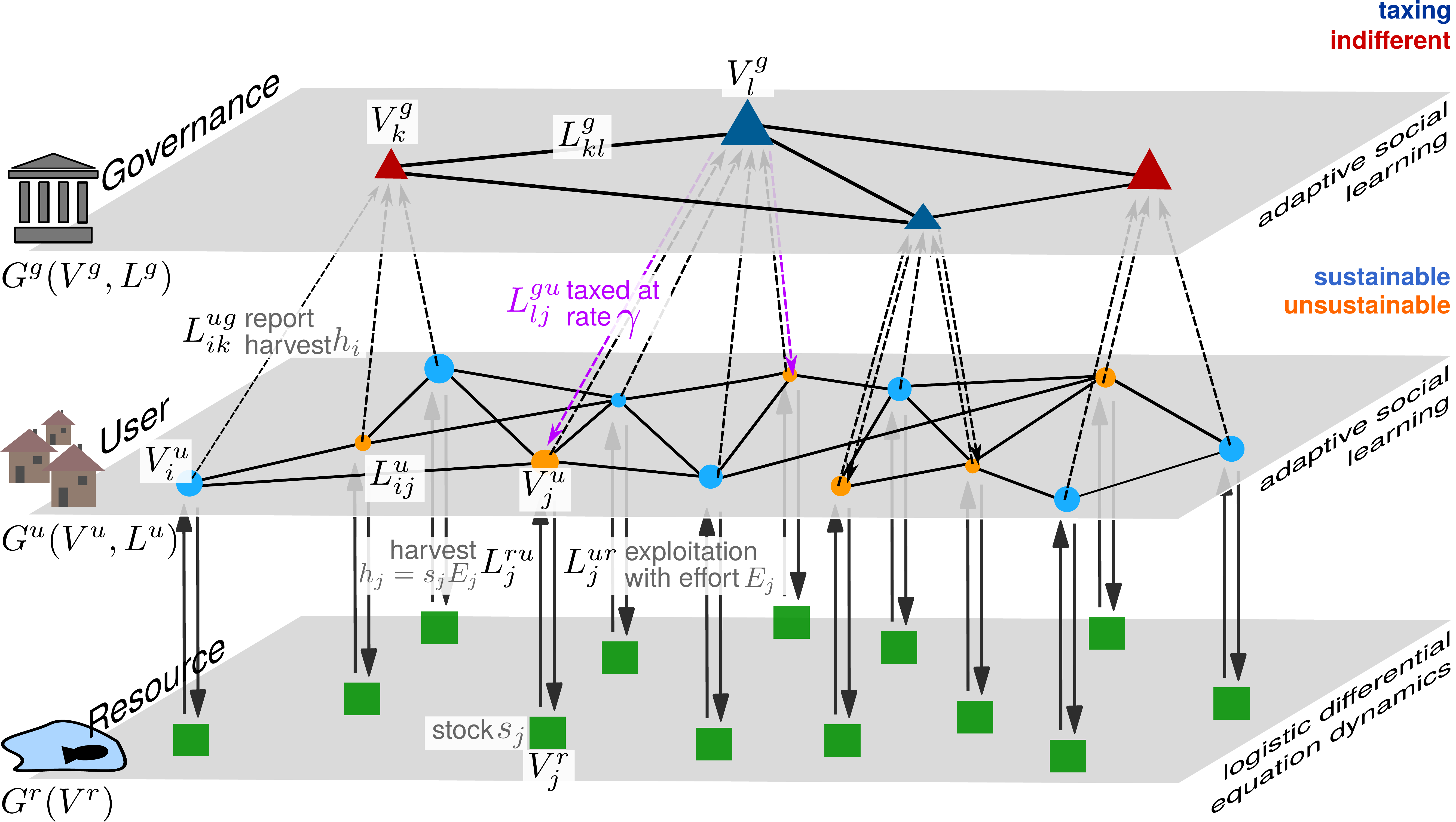}
    \caption{Schematic visualization of the interdependent three-layer model consisting of a resource layer $\resourcenetwork{}$, a user layer $\usernetwork{}$ and a governance layer $\governancenetwork{}$. 
    }\label{fig:model_scheme}
\end{figure}

In the following we describe the multilayer adaptive network model of three layers
that is studied in this work. It consists of an ecological \textit{resource layer}
representing a set of logistically growing resources, a \textit{user layer}
representing a set of agents harvesting these resources, and a
\textit{governance} layer representing agents superordinate to agents in the
user layer. Nodes in the governance layer can enforce taxes on certain types of user
behavior. The general setup of the model is summarized and visualized in
Fig.~\ref{fig:model_scheme}.

\subsection{Resource layer}
The first layer $\resourcenetwork(\resourcenodes)$ consists of a set $\resourcenodes$ of
$\numberofresourcenodes$ mutually disconnected nodes $\resourcei,\
i=1,\ldots,\numberofresourcenodes$ each representing dynamics following logistic
growth,
\begin{align}
	\frac{ds_i}{dt} = a s_i \left(1- \frac{s_i}{K}\right),
    \label{eqn:logistiGrowthundisturbed}
\end{align}
where $a$ denotes the growth rate and $K$ denotes the maximum capacity. We use identical $a$ and $K$ for all resource nodes for simplicity in this study, but heterogeneities in these properties can yield interesting effects as well~\cite{barfuss_sustainable_2017}.
Note that from here on, we use $s_i$ to denote the current state of the
resource but to also refer to the corresponding node in $\resourcenetwork$.
Without loss of generality we can set $a=K=1$ and, hence,
express the time $t$ in terms of the inverse growth rate $1/a$, and the resource stock $s_i$
in terms of the maximum capacity $K$. If undisturbed, $s_i$ displays two fixed
points $s_{i,0} = 0$ (unstable) and $s_{i,0} = 1$ (stable).

\subsection{User layer}
The second layer in our interdependent model $\usernetwork(\usernodes, \userinternallinks)$
represents a set $\usernodes$ of $\numberofusernodes=\numberofresourcenodes$
agents $i=1,\ldots,\numberofusernodes$ that harvest exactly one of the
resources $s_i$ with some effort (or strategy) $\efforti$ along a link $(i,
\resourcei)\in\extractionlinks$, where $\extractionlinks$ denotes the set of
directed links pointing from $\usernetwork$ to $\resourcenetwork$. Depending on
the currently employed effort level, a node $i\in\usernodes$ gains an
instantaneous harvest $h_i=qE_is_i$
from the resource. Here, $q$ denotes the so-called catch coefficient or efficiency
\cite{perman_natural_2003}. As we are only interested in an inter comparison of efforts across agents, we measure the effort in units of that efficiency by setting $q=1$~\cite{wiedermann_macroscopic_2015}. Harvesting
effectively reduces the amount of available stock $s_i$ to each node $i$ and
hence, Eq.~\eqref{eqn:logistiGrowthundisturbed} is adjusted to ultimately read
\begin{align}
	\frac{ds_i}{dt} = s_i \left(1- s_i\right) - E_is_i.
    \label{eqn:logistiGrowthdisturbed}
\end{align}
Each agent/node chooses between two values of
effort level $E_-=1-\Delta E$ and $E_+=1+\Delta E$ that cause the resource to
either converge into a stable fixed point $s_{i,0} = \Delta E$ for $E_-$ and
$s_{i,0}=0$ for $E_+$. We therefore denote $E_-$ the \textit{sustainable} and
$E_+$ the  \textit{unsustainable} effort. In order to further reduce the number of free parameters we set $\Delta E=0.5$
according to earlier studies
~\cite{wiedermann_macroscopic_2015,barfuss_sustainable_2017} ensuring that at the fixed point $s^*=\Delta E$
the equilibrium harvest $h_0=\Delta E(1-\Delta E)$ is maximized. 

Pairs of nodes $i\in\usernodes$ interact and update their strategies similarly
to the adaptive voter model~\cite{holme_nonequilibrium_2006,gross_adaptive_2008,gross2009adaptive,amato2017opinion,min2019multilayer} with the process
of pure imitation replaced by social
learning~\cite{traulsen_pairwise_2007,traulsen_human_2010,wiedermann_macroscopic_2015,barfuss_sustainable_2017}. Therefore, edges
$\userinternallinks$ in the user layer $\usernetwork$ indicate
a connection (such as friendship or business relationships) between the
nodes $i$ along which opinion formation takes place via the exchange of
information on current harvesting strategies $\efforti$ and corresponding harvest
$\harvesti$. To combine discrete opinion formation with continuous resource
dynamics, each node is assigned a unique
waiting time $\userwaitingtimei$ according to a Poissonian distribution that is drawn
randomly after each interaction of that corresponding node $i$,
\begin{align}
  P(\userwaitingtimei) = \frac{1}{\Delta\userwaitingtime}
  \exp\left(-\frac{\userwaitingtimei}{\Delta\userwaitingtime}\right).
    \label{eqn:probWaitingTime}
\end{align} 
Here, $\Delta\userwaitingtime$ is understood as the average waiting time of nodes in the user
layer. It directly relates to the rate of interaction between the agents as compared to the typical timescale of the resource dynamics. In that sense, a short waiting waiting time corresponds to more impatient agents while a high waiting time indicates comparatively patient agents.  

In each time step, the node $i$ with the smallest waiting time
$\userwaitingtimei$
becomes active and all stocks $s_i$ are integrated forward by
$\userwaitingtimei$. Then, a
random neighbor $j$ of $i$ is chosen such that $(i,j)\in\userinternallinks$. If
the effort levels, i.e., strategies, $\efforti$ and $\effortj$ differ, there is
a probability $\phi$ for $i$ to break its connection with $j$ and
homophilically establish a new link to a formerly unconnected node $n$ such
that $E_i=E_n$. In addition, with probability $1-\phi$, $i$ mimics the harvest
strategy of $j$ with a probability $P(E_i\rightarrow E_j)$ depending on the
difference in immediate harvest $h_i$ and $h_j$, 
\begin{equation}
	P(E_i \rightarrow E_j) = \frac{1}{2}\left(\tanh\left(h_j-h_i\right) +
    1\right).
  \label{eqn:imitation}
\end{equation}

The hyperbolic function represents the monotonic increase of the likelihood for social learning with an increase in the expected harvest differences~\cite{traulsen_pairwise_2007,traulsen_human_2010}.
After finishing one step, a new waiting time for $i$ is drawn according to
Eq.~\eqref{eqn:probWaitingTime} and added to the current $\userwaitingtimei$. This iteration
scheme continues until the model reaches a consensus state where either all
nodes in $\usernetwork$ follow the same strategy $E_i=E_j\ \forall\ i,j=1,\ldots
\numberofusernodes$ or $\usernetwork$ has fragmented into disconnected components
consisting solely of nodes with the same strategy.  
Overall, the user and resource layers follow the same dynamics as encoded in the EXPLOIT model~\cite{wiedermann_macroscopic_2015,barfuss_sustainable_2017}, given that the governance layer is in an indifferent state and, hence, exerts no influence on resource users (see below).

\subsection{Governance layer}\label{sec:governancelayer}
Social systems often obey a hierarchical structure~\cite{parsons2007outline}
including, e.g., super- and subordinate agents. To incorporate such effects, our
model additionally consists of a third layer
$\governancenetwork(\governancenodes, \governanceninternallinks)$ which, for
the sake of illustration, is denoted the \textit{governance} layer. This layer consists
of $\numberofgovernancenodes$ nodes $k$ that are connected via a set of links
$\governanceninternallinks$ indicating an abstract form of, e.g, diplomatic
relationships. Nodes $k$ can be in one of either two states
$\governancestate_k$: $\governancestate_-$ (\textit{taxing}) or $\governancestate_+$ (\textit{indifferent}), which are to some extent analogous to the sustainable and
unsustainable states of nodes in the user layer $\usernetwork$.
Additionally each node $i\in\usernodes$ in the user layer is connected to
exactly one node $k\in\governancenodes$ in the governance layer (implying that
$\numberofgovernancenodes\leq\numberofusernodes$).

Nodes $k\in\governancenodes$ also follow an opinion formation process
along the lines of the extended adaptive voter model as described above. Hence, for each node
$k\in\governancenodes$ we draw waiting times $\governancewaitingtimei$ according to
Eq.~\ref{eqn:probWaitingTime} and set an average waiting time
$\Delta\governancewaitingtime$ unique to
the governance layer $\governancenetwork$. As above, once node $k$ becomes active, a
neighbor $l$ that is connected with $k$ is drawn uniformly at random. With probability $\phi$ and if the states of
the two nodes differ (i.e., $\governancestate_l\neq \governancestate_k$), $k$ breaks its connection with $l$
and establishes a new link to a previously
unconnected node $n\in\governancenodes$, such that $\governancestate_k =
\governancestate_n$. For the sake of reducing the
number of free parameters, we
employ the same rewiring probability $\phi$ in $\usernetwork$ and $\governancenetwork$.
In contrast to nodes $i\in\usernodes$, a node $k\in\governancenodes$ does not harvest from its own resource
stock, but instead measures the cumulative harvests $h_i$ of all nodes $i$
that $k$ is connected to via interdependence links $\reportinglinks$, such
that 
\begin{align}
  h_k = \sum_{i\in\usernodes|(i,k)\in\reportinglinks} h_i,\
  k\in\governancenodes
\end{align}
Hence, the probability for a node $k\in\governancenetwork$ to update its state
$\governancestate_k$ to the state $\governancestate_l$ of one of its
neighboring nodes $l$ then reads,
\begin{align}
  P(\governancestate_k \rightarrow \governancestate_l) =
  \frac{1}{2}\left(\tanh\left(h_l-h_k\right) +
    1\right).
\end{align}
As in Eq.~\eqref{eqn:imitation} the hyperbolic tangent represents the experimentally observed increased likelihood for social learning as a function of the difference in cumulative harvest~\cite{traulsen_pairwise_2007,traulsen_human_2010}.
If $k$ is now in the taxing state, it favors the long-term sustainable
strategy $E_-$ and, hence, taxes those connected subordinate nodes
$i\in\usernodes$ that are employing the non-sustainable strategy $E_+$ at a rate
$\taxrate\in[0, 1]$. $\taxrate$ effectively lowers the harvests of
nodes $i$ with $E_i=E_+$ such that the probability for learning another node's
effort level  as given in
Eq.~\eqref{eqn:imitation} is modified to read
\begin{equation}
	P\left(E_i \rightarrow E_j\right) = \frac{1}{2} \tanh\left(\alpha_j h_j - \alpha_i h_i\right) + \frac{1}{2}
\end{equation}
Where $\alpha_i=(1-\taxrate)$ if $E_i=E_+$ and the superordinate node
$k\in\governancenodes$ of $i$ is in the taxing state (the same holds for node
$j$). Otherwise, we set $\alpha_i=1$ ($\alpha_j=1$). Thus, governance nodes $k$
in the taxing state punish unsustainable strategies of nodes $i$ in the user
layer.

In order to ensure that the social learning process in both layers reaches consensus
at approximately the same time $T_c$ we demand that
\begin{align}
  T_c = X_g \Delta \governancewaitingtime = X_u \Delta \userwaitingtime,
\end{align}
where $X_g$ and $X_u$ is the total number of pairwise interactions between
nodes in $\governancenetwork$ and $\usernetwork$, respectively. Assuming that
only learning, no adaptation and no interaction between the layers takes place,
$X_\bullet$ has previously been analytically derived as $X_\bullet=N\mu_1^2 /
\mu_2$~\cite{sood_voter_2005}, where $\mu_n$ is the n-th moment of the degree
distribution and $N$ is the number of nodes in the respective network. As we initialize
each layer as an \ER random graph~\cite{erdos_evolution_1960} with linking probability $\rho$ (see Sec.~\ref{sec:setup}) we obtain a Poissonian degree distribution with $\mu_1=\mu_2=N\rho$. This yields
\begin{align}
  &\numberofgovernancenodes^2 \rho  \Delta\governancewaitingtime = \numberofusernodes^2\rho \Delta\userwaitingtime
  \rightarrow \Delta\governancewaitingtime = (\numberofusernodes / \numberofgovernancenodes)^2 \Delta\userwaitingtime.\label{eqn:estimation}
\end{align}
Hence, we can express the average waiting time $\Delta\governancewaitingtime$ for nodes $k\in\governancenodes$
in the governance layer in terms of the average waiting time
$\Delta\userwaitingtime$ for nodes $i\in\usernodes$ in the user layer $\usernetwork$. This assumption also holds if one considers an adaptive network were only rewiring and no change in node state ($\phi=1$) takes place. Then the time to reach a fragmented state depends linearly on the number of edges that connect nodes of different states. If one considers a random network topology with only two uniformly distributed node states this number of cross-links (and thus $X_\bullet$) again depends quadratically on the number of nodes such that Eq.~\eqref{eqn:estimation} also holds for this limiting case. A further in-depth investigation of $X_\bullet$ for cases of $\phi\in(0,1)$ is beyond the scope of this work. However, for the purpose of dimension reduction within this study we assume Eq.~\eqref{eqn:estimation} to approximately hold for those cases as well.

Also note,
that since we demanded $\numberofgovernancenodes\leq\numberofusernodes$ it
follows that $\Delta\governancewaitingtime\geq\Delta\userwaitingtime$,
which is consistent with the association of network layers to users and governance actors such that governance processes commonly happen on a slower timescale than economic resource use decisions.

\subsection{Initial conditions and model setup}\label{sec:setup}

For the following analysis, we initialize our model as three coupled \ER random
graphs with $\numberofusernodes=\numberofresourcenodes=500$, 
$\numberofgovernancenodes=50$, linking probability $\rho_g=\rho_u=0.05$ for the
governance ($\governancenetwork$) and the user layer ($\usernetwork$), and linking probability $\rho_r=0$ for the
resource layer ($\resourcenetwork$). Each node $i\in\usernodes$ in the user layer is connected to
exactly one randomly drawn node $k\in\governancenodes$ in the governance layer.
All stocks in the resource layer are initially set to $s_i(t=0)=1,\ \forall\ i
=1,\ldots,\numberofresourcenodes$. For each node in the user layer, an initial
effort of $E_+$ or $E_-$ is drawn uniformly at random. The same holds (if not specified otherwise) for the
initial states of nodes in the governance layer. For each combination of the
taxrate $\taxrate$, rewiring probability $\phi$, and average waiting time in
the user layer $\Delta\userwaitingtime$, we perform Monte-Carlo simulations with $M=100$ ensemble
members until at least the user layer reaches its consensus state.

\section{Results and Discussion}
\label{sec:results}

\subsection{Social learning in the user layer}\label{sec:social_learning_user_layer}
\begin{figure}[t]
    \includegraphics[width=.75\linewidth]{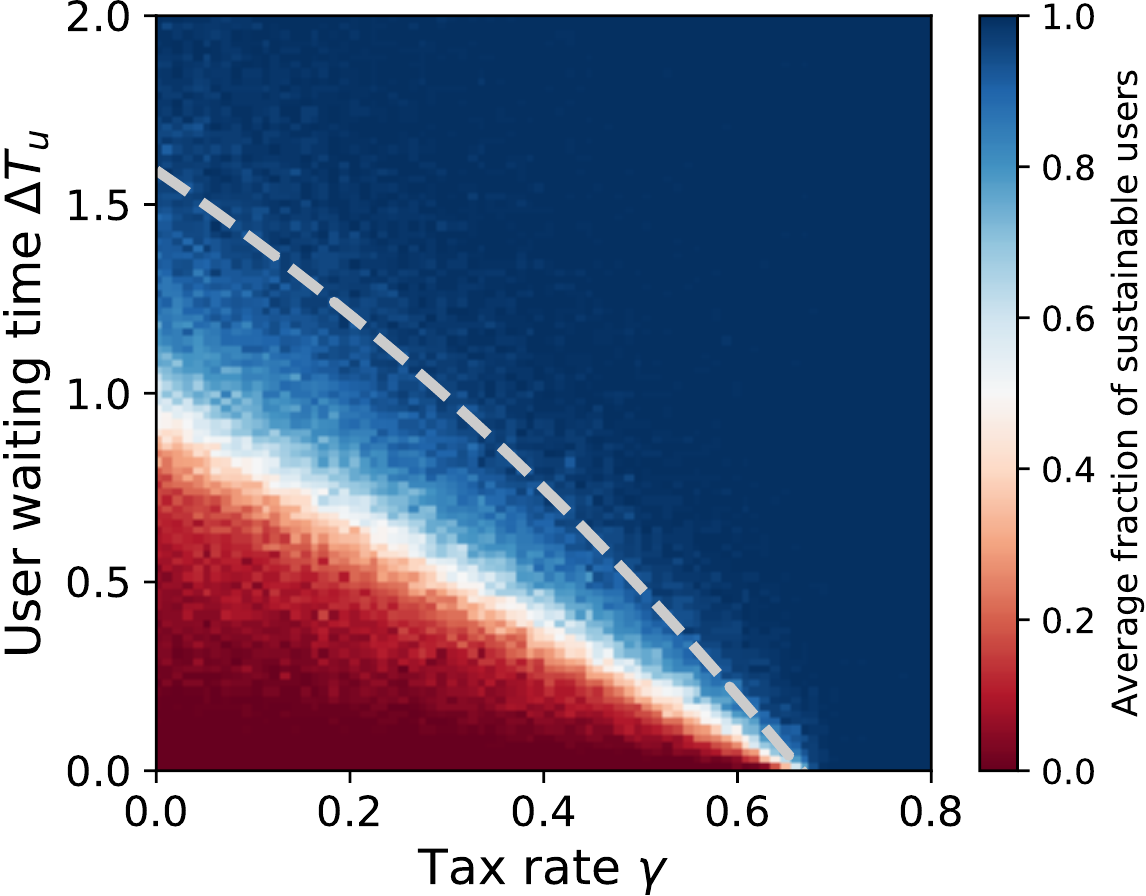}
    \caption{Average fraction of sustainable users when unsustainable nodes are always taxed at rate $\gamma$ and no rewiring takes place ($\phi=0$) The dashed line gives the  macroscopic critical Time $\tcrit{}$ as given in Eq.~\eqref{equ:criticaltimecondition2} $\tcrit{}$ decreases with increasing $\taxrate{}$ such that for a sufficiently large taxation the critical update time approaches zero at $\gammacrit{}$.}
    \label{fig:fig2_eco_dictator}
\end{figure}
We start the analysis by considering a governance network $\governancenetwork$ with only one node that is in the taxing state. This means that effectively no learning dynamics take place in the governance layer and all nodes in the user layer that employ the unsustainable strategy (effort level $E_+$) are automatically taxed at rate $\taxrate{}$. We refer to this setup as an ``eco-dictatorship" in the following.
Additionally, we first focus on the case with no adaptation in either layer and hence, set $\phi=0$. Thus, we focus on a case with solely social learning, i.e., an imitation of harvesting strategies, in the user layer. The corresponding average fraction of sustainable users in the consensus state depending on the choice of tax rate $\taxrate{}$ and user waiting time $\Delta\userwaitingtime{}$ is displayed in Fig.~\ref{fig:fig2_eco_dictator}. We mainly find that, for low tax rates $\taxrate{}$ and low user waiting times $\Delta\userwaitingtime$ the system is most likely to converge into a consensus state with all nodes employing the unsustainable strategy (lower left corner of Fig.~\ref{fig:fig2_eco_dictator}). 
This is caused by the fact that for low values of $\Delta\userwaitingtime$ most pairwise interactions take place before the resource stocks of the unsustainable agents are depleted to a state where they yield less harvest than those stocks of sustainable agents. 
In other words, for low $\Delta\userwaitingtime$ the interaction time scale becomes much shorter than the time scale of resource dynamics. With increasing tax rate $\taxrate{}$ 
the system converges more likely into a sustainable state even at comparatively low user waiting times $\Delta\userwaitingtime$ as the effective harvest of unsustainable agents is reduced more drastically. 
For very large tax rates $\taxrate{}$ and/or very large user waiting times $\Delta\userwaitingtime{}$ the system converges into a sustainable state as the resource stocks of unsustainable agents are close to their stable fixed point at $s^*=0$ when most pairwise interactions happen. On the other hand, high tax rates $\taxrate{}$ further decrease the effective harvest of unsustainable agents such that an imitation of the unsustainable strategy becomes less and less likely (Fig.~\ref{fig:fig2_eco_dictator}). 
We additionally observe that there exists a critical user waiting time $\tcrit$ above which the system always converges into a sustainable state regardless the choice of tax rate $\taxrate{}$. The same holds for $\taxrate{}$ itself as there seems to exists a critical value $\gammacrit$ above which the system also very likely converges into a sustainable state. In the following we aim to estimate values of $\tcrit$ and $\gammacrit$.

\subsection{Analytical treatment of limiting cases}

For the eco-dictatorship setup studied above, we approximate a critical update time $\tcrit$ at which the sustainable strategy $E_-$ becomes profitable in terms of immediate harvest when compared to the unsustainable one ($E_+$). For this we assume that agents do not update their strategy at times $t<\tcrit$ and hence $E_i(t)=E_i(0)\ \forall\ t<\tcrit$. In this case the temporal evolution of the corresponding stocks is obtained by integrating Eq.~\eqref{eqn:logistiGrowthdisturbed} which yields:
\begin{align}
s^{\pm}\left(t\right) = \frac{\mp\Delta E} {(\mp \Delta
    E -1 ) e^{\pm\Delta E t} + 1}\label{eqn:analytical}
\end{align}
Here $s^+(t)$ ($s^-(t)$) denotes the stock for those agents that employ the unsustainable (sustainable) strategy. If no interactions take place the two strategies yield the same harvest $h_\bullet(\tcrit{})$ at time $t=\tcrit$ and, hence,  
\begin{equation}
	(1-\gamma) s^+(\tcrit)(1+\Delta E)= s^-(\tcrit) (1-\Delta E).
    \label{equ:criticaltimecondition}
\end{equation}
Plugging in Eq.~\eqref{eqn:analytical} yields
\begin{equation}
	(1-\gamma) e^{-\Delta E \tcrit} + e^{\Delta E \tcrit} = \frac{2-\gamma-\gamma\Delta E}{1 - \Delta E^2}. \label{equ:criticaltimecondition2}
\end{equation}
This equation of the general form $a e^{-x} + e^x = b$ is solved by using $x=\ln\left(\frac{1}{2}\left(b\pm \sqrt{b^2 - 4a}\right)\right)$. Figure~\ref{fig:fig2_eco_dictator} shows the critical user waiting time $\tcrit$ as a function of the tax rate $\gamma$. We find that the approximated functional form of $\tcrit(\taxrate{})$ provides an upper bound above which the sustainable strategy almost always succeeds. As expected $\tcrit$ approaches zero with increasing $\gamma$ as higher tax rates reduce the effective harvest of unsustainable agents and, hence, makes the sustainable harvest profitable much earlier in time.

The critical tax rate $\gammacrit{}$, beyond which the sustainable resource use is always maintained (for $\gamma \geq \gammacrit{}$), can be derived by setting $\tcrit{}=0$ in Eq.~\eqref{equ:criticaltimecondition2}. This yields
\begin{equation}
    \gammacrit(\Delta E) = 2 \frac{\Delta E}{1 + \Delta E}
\end{equation}
and $\gammacrit{} = \frac{2}{3}$ for $\Delta E = 0.5$. This matches the numerically computed result of $\gammacrit \approx 0.67$ (Fig.~\ref{fig:fig2_eco_dictator}, intersection of dashed grey line with the abscissa $\Delta\userwaitingtime=0$).

\subsection{Social learning in governance and user layer }\label{sec:full_model}
\begin{figure}[t]
    \includegraphics[width=.75\linewidth]{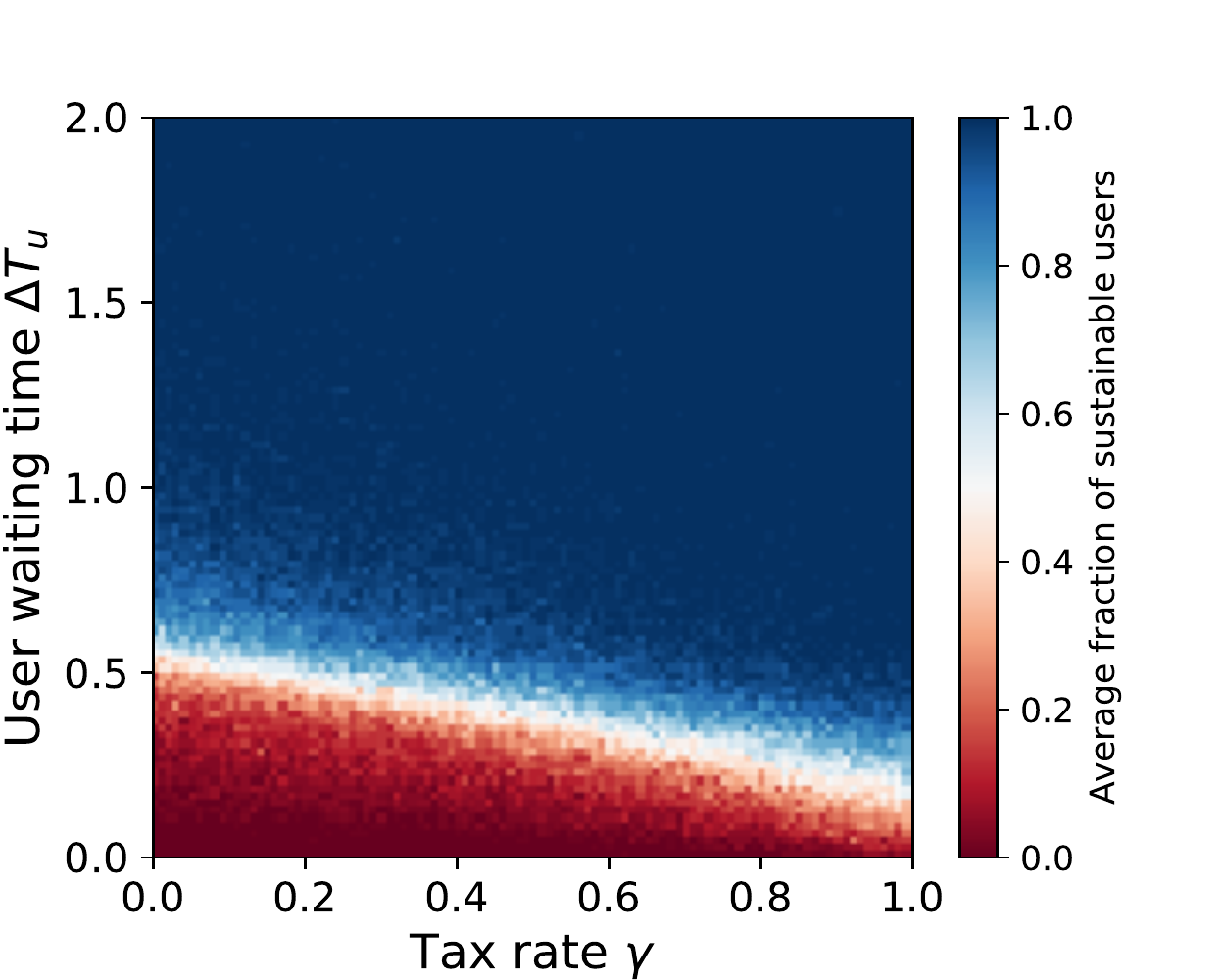}
    \caption{If social learning happens in both, the user and the governance layer, rewiring at an intermediate probability (here with $\phi=0.4$) induces a trade off. In particular, the system is more likely to become sustainable at low tax rates  but less likely to become sustainable at high tax rates as compared to a setting where unsustainable nodes are always taxed (compare Fig.~\ref{fig:fig2_eco_dictator})}. 
    \label{fig:fig3_fullmodel}
\end{figure}

After obtaining the results for a simplified governance layer with just one single node (the eco-dictatorship), we now turn to the analysis of the full model by setting $\numberofgovernancenodes=50$ and, hence, allowing for social learning as described in Sec.~\ref{sec:governancelayer} in the governance layer as well. Additionally we allow for adaptive rewiring by setting the rewiring probability to an intermediate value of $\phi=0.4$ (Fig.~\ref{fig:fig3_fullmodel}). Recall, that this implies that whenever two nodes of different state or strategy in either layer interact there is a probability of $\phi$ for the link between those two nodes to be homophilically rewired such that two nodes of the same strategy or state are connected afterwards. 

As a first general observation, we observe that increasing the tax rates $\taxrate{}$ increases the size of the sustainable regime, i.e., the system converges to a state with all nodes employing the sustainable strategy at smaller user waiting times $\userwaitingtime{}$ (Fig.~\ref{fig:fig3_fullmodel}). Hence, increasing the tax rate also increases the resilience of the entire system. 

We further find that social learning in the governance layer causes the absence of a critical tax rate $\gammacrit{}$ and, hence, there is no $\taxrate{}$ for which the system always converges into the sustainable state (compare Fig.~\ref{fig:fig2_eco_dictator} and ~\ref{fig:fig3_fullmodel}). 

At the same time we find that social learning and network adaptation induce a trade off (as compared to the case of a single sustainable governance node) where the sustainable regime increases in size for the case of small tax rates $\taxrate{}$ but decreases in size for larger tax rates $\taxrate{}$ (compare again the size of the sustainable regimes between Fig.~\ref{fig:fig2_eco_dictator} and ~\ref{fig:fig3_fullmodel}). This phenomenon is explained in the following.

Since the governance layer now partly consists of nodes that are in the \textit{indifferent} state, there is a chance for unsustainable nodes in the user layer to not being taxed. In that case, their harvest likely exceeds that of sustainable nodes in the beginning of the simulation if the average user waiting time $\Delta\userwaitingtime$ is small. At the same time, even if unsustainable nodes are being taxed at a moderate rate their harvest might exceed that of sustainable nodes if their corresponding stocks are still far away from equilibrium, i.e., depletion. Hence, the increased size of the unsustainable regime at larger tax rates can be attributed to the effect of social learning in the governance layer. 

For low tax rates we observe a decrease in size of the unsustainable regime as compared to the case of no social learning in the governance layer (Sec.~\ref{sec:social_learning_user_layer}) and, more importantly, no network adaptation in either layer. It has been observed already in earlier studies that adaptation fosters the tendency of the system to reach the sustainable state as it allows nodes of the same strategy to form clusters~\cite{wiedermann_macroscopic_2015}. This clustering of specifically the sustainable nodes allows them to avoid exposure to the unsustainable strategy ($E_+$) until the sustainable strategy ($E_-$) has become more profitable. From there on the sustainable strategy can spread through the network and tip the entire system into a sustainable state even at lower user waiting times as compared to the case with no adaptation (compare lower left parts of Fig.~\ref{fig:fig2_eco_dictator} and ~\ref{fig:fig3_fullmodel}). Hence, the decrease in the size of the unsustainable regime for lower tax rates is mainly attributed to the presence of adaptation in both, the user and the governance layer.

In summary, we find that social learning in the governance layer increases the size of the unsustainable regime at high tax rates when compared to the case of an absence of social learning. At the same time, adaptive rewiring increases the size of the sustainable regime at lower tax rates. In other words, at low tax rates network adaptation and governmental social learning are preferred to drive the system into a sustainable state, while at high tax rates social learning and adaptation are to be avoided.

\subsection{Comprehensive analysis}
\begin{figure}[t]
\centering
\includegraphics[width=\textwidth]{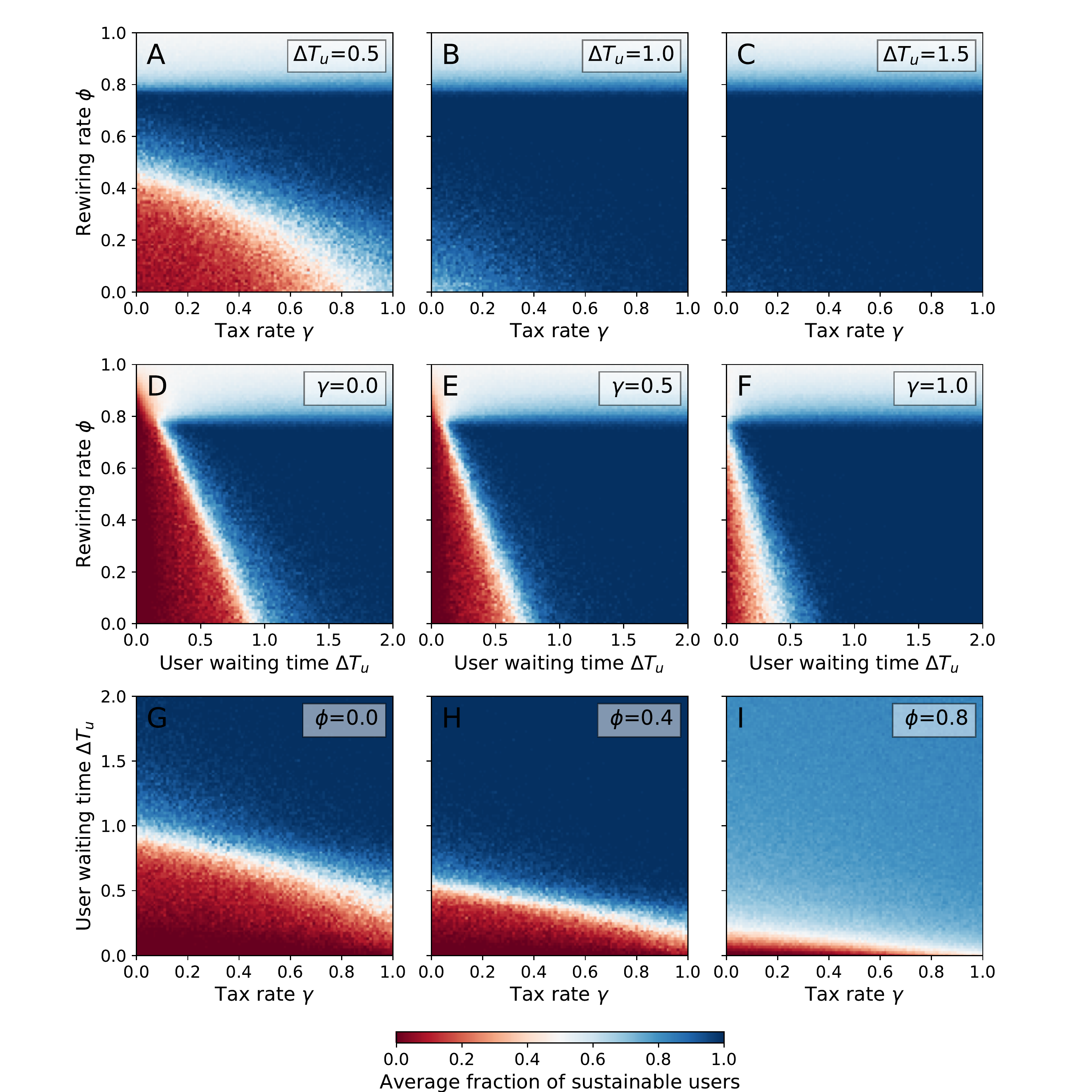}
\caption{The average fraction of sustainable agents for different combinations of tax rate $\taxrate{}$, user waiting time $\Delta\userwaitingtime$ and adaptation probability $\phi$.}
\label{fig:fig4_comprehensive}
\end{figure}

We ultimately vary the three crucial parameters $\taxrate{}$ (tax rate), $\phi$ (adaptation/rewiring probability) and $\Delta\userwaitingtime$ (average user waiting time) to provide a comprehensive analysis and to illustrate how the size of the sustainable regime depends on their particular choices (Fig.~\ref{fig:fig4_comprehensive}). We summarize our three main results below:

i.) First, we present results for three choices of average user waiting time $\Delta \userwaitingtime=0.5$ (Fig.~\ref{fig:fig4_comprehensive}A), $\Delta \userwaitingtime=1.0$ (Fig.~\ref{fig:fig4_comprehensive}B) and $\Delta \userwaitingtime=1.5$ (Fig.~\ref{fig:fig4_comprehensive}C) and varying values of $\taxrate{}$ and $\phi$. For all three cases, we first observe that there exists a fragmentation threshold at around $\phi\approx0.8$ above which the final share of sustainable nodes in the user layer roughly corresponds to the expected initial share of 0.5 (Fig.~\ref{fig:fig4_comprehensive}A-C). In addition we find that for $\Delta \userwaitingtime=0.5$ there exists an unsustainable regime for low tax rates $\taxrate{}$ and low rewiring probabilities $\phi$ as the myopic agents do not foresee a potential collapse of their respective resource stocks when being unsustainable or indifferent. However, the size of this regime decreases in size with increasing $\taxrate{}$ (Fig.~\ref{fig:fig4_comprehensive}A) as the unsustainable strategy becomes less profitable. With increasing the user waiting time to $\Delta \userwaitingtime=1.0$ (Fig.~\ref{fig:fig4_comprehensive}B) or $\Delta \userwaitingtime=1.5$ (Fig.~\ref{fig:fig4_comprehensive}C) the system converges into a sustainable state for almost all choices of $\taxrate{}$ and $\phi$ as long as the rewiring rate is chosen such that the fragmentation threshold is not transgressed. Hence, we conclude that the larger the average user waiting time, the more likely the system converges into a sustainable state (as it is also reported in earlier studies~\cite{wiedermann_macroscopic_2015,barfuss_sustainable_2017}). 

ii.) Next, we present the results for three choices of tax rate $\taxrate{}=0$ (no taxation), $\taxrate{}=0.5$ (intermediate taxation) and $\taxrate{}=1$ (full taxation)  and varying user waiting time $\Delta\userwaitingtime$ as well as rewiring probability $\phi$ (Fig.~\ref{fig:fig4_comprehensive}D-F). For the case of no taxation, i.e., no effect of the governance layer, (Fig.~\ref{fig:fig4_comprehensive}D) the size of the sustainable regime increases linearly with increasing $\phi$ until, again, the fragmentation transition is reached. This result is in accordance with earlier studies that investigate the effect of social learning and adaptation in a system that is only comprised of the user and the resource layer~\cite{wiedermann_macroscopic_2015}. Increasing the tax rate steadily decreases the size of the unsustainable regime (Fig.~\ref{fig:fig4_comprehensive}E+F), while the linear dependence between the rewiring probability $\phi$ and the size of the regime sizes persists. Remarkably, even for the case of full taxation (Fig.~\ref{fig:fig4_comprehensive}F) the unsustainable regime remains to exist as for very low user waiting times $\Delta\userwaitingtime$ the unsustainable and indifferent strategies can spread through both layers as the resource stocks deplete slower as compared to the rate of social interactions. 

iii.) Ultimately, we consider three cases of rewiring probabilities, i.e.,  $\phi=0$ (no rewiring and only social learning), $\phi=0.4$ (intermediate rewiring) and $\phi=0.8$ (almost only rewiring at a rate close to the fragmentation threshold and few cases of social learning), Fig.~\ref{fig:fig4_comprehensive}G-I. Note that Fig.~\ref{fig:fig4_comprehensive}H shows the same results as the previously discussed Fig.~\ref{fig:fig3_fullmodel}. As already discussed in Sec.~\ref{sec:full_model} social learning in the governance layer causes the absence of a critical tax rate $\gammacrit{}$, that we observed from the case of a single sustainable governance node, Fig.~\ref{fig:fig4_comprehensive}G. However, allowing for rewiring at an intermediate rate (Fig.~\ref{fig:fig4_comprehensive}H) again yields an increase in the size of the sustainable regime. Further increasing the rewiring probability causes the size of the sustainable regime to increase even further. However, as the system approaches the fragmentation transition, the average fraction of sustainable users is lowered due to the formation of isolated clusters of user and governance nodes that solely employ the unsustainable/indifferent strategy (Fig.~\ref{fig:fig4_comprehensive}I).

In summary, we observe that the system is most likely to reach a sustainable regime if a high tax rate $\taxrate{}$ and a rewiring probability $\phi$ close to (but still below) the fragmentation transition are chosen. In other words, such a combination of parameters maximizes the size of the sustainable regime. Given that an implementation of arbitrarily high tax rates is often not feasible, minimal/optimal tax rates could be chosen for a given user waiting time $\Delta \userwaitingtime$ and rewiring probability $\phi$ such that the system is likely to converge into a sustainable state while putting the least amount of pressure as possible on to users that show an undesired strategy (see e.g., Fig.~\ref{fig:fig4_comprehensive}A).

\section{Conclusion}
\label{sec:conclusions}

In this article, we have developed a stylized model for polycentric hierarchical governance structures with a focus on investigating the preconditions for the sustainable use of renewable resources. While resource users can employ either a sustainable or non-sustainable harvesting strategy, policies are implemented via either taxation or no taxation of non-sustainable resource use.
The model design is targeted towards a better systems understanding. Governance actors' and resource users' interactions are driven by the following two social processes: social learning of favorable strategies and homophilic network adaptation, but take place on different hierarchical scales.

Generally we find that sustainability is favored for slow interaction timescales, large homophilic network adaptation (as long it is below the fragmentation threshold) and high taxation rates. 
For the case of an eco-dictatorship, where a single governance actor taxes all non-sustainable behavior, we find the intuitive result that a sufficiently large taxation rate always causes a sustainable outcome.
In contrast, in the fully process-driven model with social learning and homophilic network adaptation among governance actors, we find a trade-off: sustainability is enhanced for low and hindered for high tax rates compared to the results obtained for the eco-dictatorship.

This rather non-intuitive result highlights that the emergent outcomes of freely co-evolving social processes can be preferable compared to those obtained with a benevolent centralistic actor, if low tax rates are a normative preference. 
In this regard, our model serves as a stylized example to find minimal tax rates that still guarantee an optimally sustainable outcome given polycentric governance structures, given a social learning process with a certain network adaptation rate and interaction timescale.

Possibly, our model could serve as a prototype for more detailed studies to be targeted at the question of optimal carbon taxes rates \cite{klenert2018making}.
It highlights how social processes such as opinion formation may be combined with macro-economic optimization techniques \cite{donges2018earth} in order to gain momentum on the road to the much needed rapid decarbonization \cite{rockstrom2017roadmap}.

\begin{acknowledgement}
This work was conducted in the framework of the COPAN collaboration (www.pik-potsdam.de/copan) at the Potsdam Institute for Climate Impact Research (PIK). We are grateful for financial support by the Heinrich B\"oll Foundation, Leibniz Association (project DominoES), Stordalen Foundation (via the Planetary Boundary Research Network PB.net), the Earth League's EarthDoc program, and the European Research Council (ERC advanced grant project ERA). Parameter studies were performed on the high-performance compute cluster of PIK, supported by the European Regional Development Fund, BMBF, and the Land Brandenburg. The authors thank Jobst Heitzig and Wolfgang Lucht for inspiring discussions.
\end{acknowledgement}


\begin{thebibliography}{}

\bibitem{arneth2014global}
 A Arneth, C Brown, MDA Rounsevell, Nature Climate Change \textbf{4}, (2014) 550 7
\bibitem{traulsen_human_2010}
 Arne Traulsen, Dirk Semmann, Ralf D Sommerfeld et al., Proc. Natl. Acad. Sci. U.S.A. \textbf{107}, (2010) 2962--2966 7
\bibitem{traulsen_pairwise_2007}
 Arne Traulsen, Jorge M. Pacheco, Martin A. Nowak, Journal of Theoretical Biology \textbf{246}, (2007) 522--529 3
\bibitem{min2019multilayer}
 Byungjoon Min, Maxi San Miguel, New Journal of Physics \textbf{21}, (2019) 035004 
\bibitem{schleussner2016clustered}
 Carl-Friedrich Schleussner, Jonathan F Donges, Denis A Engemann et al., Scientific Reports \textbf{6}, (2016) 30790 
\bibitem{herrmann2018case}
 Carsten Herrmann-Pillath, Ecological Economics \textbf{149}, (2018) 212--225 
\bibitem{koliba2018governance}
 Christopher J Koliba, Jack W Meek, Asim Zia et al., \textit{Governance networks in public administration and public policy} (Routledge, 2018)
\bibitem{huepe2011adaptive}
 Cristi{\'a}n Huepe, Gerd Zschaler, Anne-Ly Do et al., New Journal of Physics \textbf{13}, (2011) 073022 7
\bibitem{klenert2018making}
 David Klenert, Linus Mattauch, Emmanuel Combet et al., Nature Climate Change \textbf{8}, (2018) 669--77 8
\bibitem{mullerhansen2019can}
 Finn M{\"u}ller-Hansen, Jobst Heitzig, Jonathan F Donges et al., Ecological Economics \textbf{159}, (2019) 198--211 
\bibitem{muller2017towards}
 Finn M{\"u}ller-Hansen, Maja Schl{\"u}ter, Michael M{\"a}s et al., Earth System Dynamics \textbf{8}, (2017) 977--1007 4
\bibitem{couzin2011uninformed}
 Iain D Couzin, Christos C Ioannou, G{\"u}ven Demirel et al., Science \textbf{334}, (2011) 1578--1580 6062
\bibitem{farmer2019sensitive}
 JD Farmer, C Hepburn, MC Ives et al., Science \textbf{364}, (2019) 132--134 6436
\bibitem{mathias2018does}
 Jean-Denis Mathias, John M Anderies, Marco Janssen, Earth's Future \textbf{6}, (2018) 1555--1567 11
\bibitem{mathias2017multi}
 Jean-Denis Mathias, Steven Lade, Victor Galaz, International Journal of the Commons \textbf{11}, (2017)  1
\bibitem{heitzig2016topology}
 Jobst Heitzig, Tim Kittel, Jonathan F Donges et al., Earth System Dynamics \textbf{7}, (2016)  1
\bibitem{rockstrom2017roadmap}
 Johan Rockstr{\"o}m, Owen Gaffney, Joeri Rogelj et al., Science \textbf{355}, (2017) 1269--1271 6331
\bibitem{anderies2019knowledge}
 John M Anderies, Jean-Denis Mathias, Marco A Janssen, Proceedings of the National Academy of Sciences \textbf{116}, (2019) 5277--5284 12
\bibitem{donges2018earth}
 Jonathan F Donges, Jobst Heitzig, Wolfram Barfuss et al., Earth System Dynamics Discussions, (2018) 1--27 
\bibitem{donges2017closing}
 Jonathan F Donges, Ricarda Winkelmann, Wolfgang Lucht et al., The Anthropocene Review \textbf{4}, (2017) 151--157 2
\bibitem{renn2017extended}
 J{\"u}rgen Renn, Manfred Laubichler, \textit{Extended Evolution and the History of Knowledge} (in Integrated History and Philosophy of Science, 2017) 109--125
\bibitem{horstmeyer2018network}
 Leonhard Horstmeyer, Christian Kuehn, Stefan Thurner, Physical Review E \textbf{98}, (2018) 042313 4
\bibitem{milkoreit2018social}
 Manjana Milkoreit, Jennifer Hodbod, Jacopo Baggio et al., Environmental Research Letters \textbf{13}, (2018) 033005 3
\bibitem{wiedermann_macroscopic_2015}
 Marc Wiedermann, Jonathan F. Donges, Jobst Heitzig et al., Physical Review E \textbf{91}, (2015) 052801 5
\bibitem{erdos_evolution_1960}
 P. Erdős, A. Rényi, \textit{On the {Evolution} of {Random} {Graphs}} (in Publication of the {Mathematical} {Institute} of the {Hungarian} {Academy} of {Sciences}, 1960) 17--61
\bibitem{klamser2017zealotry}
 Pascal P Klamser, Marc Wiedermann, Jonathan F Donges et al., Physical Review E \textbf{96}, (2017) 052315 5
\bibitem{verburg2016methods}
 Peter H Verburg, John A Dearing, James G Dyke et al., Global Environmental Change \textbf{39}, (2016) 328--340 
\bibitem{holme_nonequilibrium_2006}
 Petter Holme, M. E. J. Newman, Physical Review E \textbf{74}, (2006) 056108 5
\bibitem{amato2017opinion}
 Roberta Amato, Nikos E Kouvaris, Maxi San Miguel et al., New Journal of Physics \textbf{19}, (2017) 123019 12
\bibitem{perman_natural_2003}
 Roger Perman, \textit{Natural resource and environmental economics} (Pearson Education, 2003)
\bibitem{jain1998autocatalytic}
 Sanjay Jain, Sandeep Krishna, Physical Review Letters \textbf{81}, (1998) 5684 25
\bibitem{jain2001model}
 Sanjay Jain, Sandeep Krishna, Proceedings of the National Academy of Sciences USA \textbf{98}, (2001) 543--547 2
\bibitem{boccaletti2014structure}
 Stefano Boccaletti, Ginestra Bianconi, Regino Criado et al., Physics Reports \textbf{544}, (2014) 1--122 1
\bibitem{lade2017modelling}
 Steven J Lade, {\"O}rjan Bodin, Jonathan F Donges et al., arXiv preprint arXiv:1704.06135, (2017)  
\bibitem{parsons2007outline}
 Talcott Parsons, \textit{An Outline of the Social System [1961]} (na, 2007)
\bibitem{gross_adaptive_2008}
 Thilo Gross, Bernd Blasius, J. R. Soc. Interface \textbf{5}, (2008) 259--271 20
\bibitem{gross_epidemic_2006}
 Thilo Gross, Carlos J. Dommar D'Lima, Bernd Blasius, Phys. Rev. Lett. \textbf{96}, (2006) 208701 20
\bibitem{gross2009adaptive}
 Thilo Gross, Hiroki Sayama, \textit{Adaptive networks} (in Adaptive Networks, 2009) 1--8
\bibitem{sood_voter_2005}
 V. Sood, S. Redner, Physical Review Letters \textbf{94}, (2005) 178701 17
\bibitem{barfuss_sustainable_2017}
 W. Barfuss, J. F. Donges, M. Wiedermann et al., Earth System Dynamics \textbf{8}, (2017) 255--264 2
\bibitem{steffen2018trajectories}
 Will Steffen, Johan Rockstr{\"o}m, Katherine Richardson et al., Proceedings of the National Academy of Sciences \textbf{115}, (2018) 8252--8259 33

\end{thebibliography}
\end{document}